# Chiral magnetic interlayer interaction in synthetic antiferromagnets

Dong-Soo Han[1,2,3†], Kyujoon Lee[1†], Jan-Philipp Hanke[1,4], Yuriy Mokrousov[1,4], Kyoung-Whan Kim[2], Woosuk Yoo[5], Youri Van Hees[3], Tae-Wan Kim[6], Reinoud Lavijsen[3], Chun-Yeol You[7], Henk J. M. Swagten[3], Myung-Hwa Jung[5*], and Mathias Kläui[1*]

1. Institute of Physics, Johannes Gutenberg-Universität Mainz, 55099, Mainz, Germany

2. Center for Spintronics, Korea Institute for Science and Technology, Seoul, Republic of Korea.

3. Department of Applied Physics, Institute for Photonic Integration, Eindhoven University of Technology, P.O. Box 513, 5600 MB Eindhoven, The Netherlands

4. Peter Grünberg Institut and Institute for Advanced Simulation, Forschungszentrum Jülich and JARA, 52425 Jülich, Germany

5. Department of Physics, Sogang University, Seoul, Republic of Korea.

6. Department of Advanced Materials Engineering, Sejong University, Seoul, Republic of Korea

7. Department of Emerging Materials Science, DGIST, Daegu 42988, Republic of Korea

**The exchange interaction underlies ferroic magnetic interactions and is thus the key element that governs statics and dynamics of magnetic systems. This fundamental interaction comes in two flavors -** *symmetric* **and** *antisymmetric* **interactions. While the symmetric interaction leads to ferro- and antiferromagnetism, the** *antisymmetric* **interaction has attracted significant interest owing to its major role in promoting topologically non-trivial spin textures**[1–8] **that promise high-speed and energy-efficient devices.**[1,9–11] **So far, the** *antisymmetric* **exchange interaction, which is rather short–ranged and limited to a single magnetic layer has been demonstrated**[1,12]**, while the** *symmetric* **interaction also leads to long-range** *interlayer* **exchange interaction. Here, we report the missing component of the long-range** *antisymmetric interlayer* **exchange interaction in perpendicularly magnetized synthetic antiferromagnets with parallel and antiparallel magnetization alignments. Asymmetric hysteresis loops under an in-plane field unambiguously reveal a unidirectional and chiral nature of this novel interaction, which cannot be accounted for by existing interaction mechanisms, resulting in canted magnetization alignments. This can be explained by spin-orbit coupling combined with reduced symmetry in multilayers. This new class of chiral interaction provides an additional degree of freedom for engineering magnetic structures and promises to enable a new class of three-dimensional topological structures.**[13,14]

† These authors contributed equally to this work.

* Authors to whom correspondence should be addressed: mhjung@sogang.ac.kr & klaeui@uni-mainz.de

Ferromagnets (FMs) and antiferromagnets (AFMs) possess collinear spin alignments within magnetic domains, due to an interaction, which is called *symmetric* or Heisenberg exchange interaction. While this conventional interaction is well known, recently a different interaction has moved into the forefront of interest, which leads to non-collinear and chiral spin textures. This new class of exchange interactions - *antisymmetric exchange interaction* or Dzyaloshinskii-Moriya interaction (DMI)[15–17] – stems from the spin-orbit coupled electrons (or ions), which mediate exchange interaction between neighboring spins within a FM, and an inversion symmetry breaking (ISB) resulting in a finite amplitude of the net effect.[17–19] The *antisymmetric* component of different kinds of exchange interactions has been discovered in a variety of materials systems: Examples include the *antisymmetric* terms of super-exchange and *s-d* exchange interactions in AFM insulators[15,16] and metallic spin glasses[20], respectively. Very recently, the discovery of the *intralayer* DMI, arising in systems with ISB at interfaces between ferromagnet/heavy metal layers (Fig. 1a) which leads to an *antisymmetric* interaction between the magnetic moments within the ferromagnetic layer, has stimulated work in the field of spintronics. In particular, it has opened fascinating new avenues for fundamental research[21] as well as highly efficient and fast spin-based information technologies.[1,9–11]

Beyond this *intralayer* exchange interaction, there can be an exchange interaction also across ferromagnetic layers through an indirect *interlayer* exchange interaction (IEI) *viz.,* Ruderman-Kittel-Kasuya-Yosida (RKKY) interaction. The IEI that couples the magnetic moments in two separate layers in a collinear fashion is a crucial element for many applications in modern magnetic storage devices and spin electronics as it enables synthetic antiferromagnets, and recently, has attracted a renewed interest in the community in line with an emerging field of antiferromagnetic spintronics.[10,22,23] The studies on the IEI, however, have so far focused only on its *symmetric* part. But from symmetry considerations, one expects that the IEI can also lead to the emergence of an *antisymmetric* IEI. Specifically, an *antisymmetric* IEI is possible in systems with ISB in the plane of thin films (yellow and green boxes in Fig. 1b). An interesting feature of the *antisymmetric* IEI is that it leads to 3D chiral magnetization configurations perpendicular to the film plane, in contrast to the *antisymmetric* component of the *intralayer* exchange interactions leading to chiral 2D spin structures confined only within individual magnetic layers. This opens the possibility for designing three-dimensional topological structures. Despite its fundamental importance as well as the associated technological promises,[10,14,22,24] clear experimental evidence of the *antisymmetric* IEI is conspicuously elusive so far.[25,26]

In this Letter, we present the experimental demonstration of such a hitherto uncovered *antisymmetric* IEI in perpendicularly magnetized synthetic antiferromagnets (SAFs) with parallel and antiparallel magnetization alignments. We study the multilayer reversal in different stacks and using judiciously designed field sequences, we can identify from unidirectional and chiral magnetization reversal the presence of an *antisymmetric* IEI and understand this interaction based on ab-initio calculations.

We start by developing the necessary concepts to unambiguously identify the effect of *antisymmetric* IEI. In general, the magnetization reversal in FMs is invariant upon the inversion of the magnetic field direction. However, this field-reversal invariance does not hold if the inversion symmetry is broken in a given physical system. One particular example is the *intralayer* DMI.[17] In the presence of *intralayer* DMI, domain walls (DWs) experience different effective fields according to their magnetic orderings, up-to-down (U-D) and down-to-up (D-U), under an in-plane magnetic field $H_{IN}$ as the core magnetizations within DWs of U-D and D-U align along opposite directions due to their preferred handedness by DMI. Consequently, when the DW moves, its dynamics becomes asymmetric with respect to $H_{IN}$, depending on their magnetic ordering.[1,3,27,28]

Analogously, the *antisymmetric* IEI can break the field-reversal symmetry for the magnetization reversal. In the absence of the *antisymmetric* IEI, $H_{IN}$ cannot break the inversion symmetry but only assist to lower the energy barrier for the magnetization reversal independent of the switching polarity (left panels of Fig.1c and 1d). However, if the *antisymmetric* IEI is present, the chiral magnetization configurations are affected differently by $H_{IN}$, assisted or hindered in their magnetization switching depending on the sign of $H_{IN}$ and the magnetization configurations. Particularly, they exhibit contrasting energy barriers for magnetization switching from parallel to antiparallel and antiparallel to parallel alignments as well as for switching of D-U and U-D, as shown in right panels of Fig. 1c and 1d (Supplementary Note 1). Accordingly, one would expect different switching fields with respect to the sweeping direction of the magnetic field, which in turn results in the asymmetric magnetic hysteresis loops.

To test experimentally if the aforementioned asymmetric switching exists, which would indicate the presence of *antisymmetric* IEI, we measure the switching fields of typical SAFs of Ta(4)/ Pt(4)/ Co(0.6)/ Pt(0.5)/ Ru($t_{Ru}$)/ Pt(0.5)/ Co(1)/ Pt(4) (layer thicknesses in nanometers), by sweeping the out-of-plane magnetic field, $H_z$, whilst simultaneously applying $H_{IN}$ (Methods section). Here two Co layers are coupled to each other via the *symmetric* IEI and perpendicularly magnetized with either parallel or antiparallel magnetization alignments at remanence. The magnetic hysteresis loops are measured by anomalous Hall effect (AHE), using the measurement configurations shown in Fig. 1e. For comparison, we also measure the switching fields of the reference sample Pt/Co/Pt/Ru that is nominally the same as the bottom half of the SAFs but without any IEI.

Figure 2a shows the magnetic hysteresis loops of Pt/Co/Pt/Ru and Pt/Co/Pt/Ru/Pt/Co/Pt where $t_{Ru}$=0.4 and 2.7 nm, for which the *symmetric* IEI is ferromagnetic and antiferromagnetic leading to parallel and antiparallel alignment of the layers, respectively. Square hysteresis loops are clearly seen for all structures, showing that they have strong perpendicular magnetic anisotropy (PMA). Importantly, we find that the hysteresis loops for the SAFs with parallel and antiparallel coupling become significantly asymmetric when $H_{IN}$ is applied. For the parallel coupling case, at $|\mu_0 H_{IN}|$ = 100mT, a difference of approximately 0.7 mT in the switching fields ($\Delta\mu_0 H_{SW}$) between U-D and D-U is found. For the antiparallel coupling case, the hysteresis loop is seemingly biased to the left (right) at $\mu_0 H_{IN}$ = 100mT (-100mT), giving rise to $\Delta\mu_0 H_{SW}$ = 1.1 and 1.4 mT for switching from parallel to antiparallel and from antiparallel to parallel alignments, respectively. Such asymmetric behavior is in striking contrast to the results obtained from our reference sample of Pt/Co/Pt/Ru, where the magnetic hysteresis loops are symmetric with respect to $H_z = 0$ irrespectively of the sign of $H_{IN}$. The measured absence of inversion symmetry in the hysteresis loops is in obvious disagreement with the field-reversal symmetry, demonstrating the presence of a symmetry-breaking interaction such as *antisymmetric* IEI in our SAFs. Moreover, we note that the field-reversal symmetry for Pt/Co/Pt/Ru in the same setup also excludes any possible artifact from the misalignment of the in-plane magnet, which could otherwise cause an asymmetry in the hysteresis loop.

To understand the origin of the asymmetric switching behavior, we next measure the azimuthal-angular dependence of $H_{SW}$, as shown in Fig. 2b and 2c. Here, the magnitude of the in-plane field is kept at $|\mu_0 H_{IN}| = 100$ mT, while rotating from 0° to 360°. In systems with inversion symmetry, one expects to see an isotropic or uniaxial (or multiaxial) anisotropy depending on the crystalline properties of thin films, which is indeed found in our reference sample (see Fig. 2b). Notably, however, we find that the magnetization switching for both SAFs with parallel and antiparallel alignment exhibits a unidirectional anisotropy which is for parallel (antiparallel) alignment with symmetric (**S**) and asymmetric (**AS**) along the direction of $\mathbf{H}_{IN}$ // 75° (150°) and $\mathbf{H}_{IN}$ // 165° (240°), respectively (this will be discussed in detail later). This highlights the *unidirectional* nature of the observed *interlayer* interaction. Interestingly, for the antiparallel coupling, we obtain markedly different unidirectional features in the two magnetic layers: for the case of the top Co layer (FM$_{top}$), the value of $|\mu_0 H_{SW}|$ for the U-D (D-U) is biased to 60° (240°), while for the bottom Co layer (FM$_{bottom}$), it is biased along the opposite direction. This opposite unidirectional behavior between two magnetic layers unambiguously reveals that the observed unidirectional effect has a chiral nature (see Supplementary Note 1) in line with an *antisymmetric* IEI. Here, we would like to note that the observed chiral behavior is radically different from that expected from currently known magnetic interactions. For example, the biquadratic IEI is also responsible for non-collinear configurations,[29] but leads to an isotropic behavior without preferred handedness, which is contrary to our observations as seen in Fig. 2c. Furthermore, the current *intralayer* DMI cannot account for such asymmetric switching behavior, because this interaction cannot produce the obtained asymmetric hysteresis unless it is combined with additional symmetry breaking effects such as DC spin currents[27] or laterally asymmetric nanostructures[28] (see Supplementary Note 2).

The *antisymmetric* IEI is expected in particular to modify the dependence of $H_{SW}$ on $H_{IN}$, which we plot in Fig. 3. For the structure with parallel coupling, the asymmetric behavior between U-D and D-U switching is again clearly found for the case where the $H_{IN}$ is applied along the **AS** axis, while almost symmetric behavior is seen for $\mathbf{H}_{IN}$ // **S**. (Fig. 3a and 3c) In particular, for the antiparallel coupling case, one can see that $\mathbf{H}_{IN}$ for local maxima (or minima) shifts away from $H_{IN} = 0$ mT for $\mathbf{H}_{IN}$ // **AS**, and the direction of the shift reverses for the opposite switching polarity. (Fig. 3b) This shift of $H_{SW}$ along the $H_{IN}$ axis is a robust indicator for the presence of the *antisymmetric* IEI; the offset in curves of $H_{SW}$ vs. $H_{IN}$ indicates the presence of a built-in effective field, the sign and magnitude of which rely on the relative orientation of the magnetization between the top and bottom Co layers. This is analogous to the internal fields from the *intralayer* DMI, which depends on the magnetic ordering of DW structures.[27] However, this is in sharp contrast to the case without the *antisymmetric* IEI, where $H_{IN}$ always assists in switching the magnetization of perpendicularly magnetized materials irrespectively of the sign of $H_{IN}$ and switching polarity.

To validate the observed asymmetric switching behavior by the *antisymmetric* IEI, we perform numerical calculations based on a macro-spin model incorporating the *symmetric* and *antisymmetric* IEI (details, see Methods section). The calculated azimuthal-angular and field-dependence of $H_{SW}$ for the parallel and antiparallel couplings are presented in Fig. 3c and 3d, respectively. While we reproduce the experimentally found asymmetric switching of the two layers, the order of the switching is found to be opposite for our numerical calculations, which is most likely due to thermal effects and imperfections in the experiment that are not incorporated into the calculations.[32] Taking into account the reversed switching sequence, however, we find that the numerical calculations are qualitatively in good agreement with the experimental data, clearly reproducing the key signatures, which are asymmetric and off-centered $H_{SW}$ vs. $H_{IN}$ as well as the unidirectional and chiral azimuthal-angular dependence of $H_{SW}$ (see Supplementary

Note 3), and these are the characteristic features that can only occur due to the presence of an *antisymmetric* IEI. From this we can conclude that the unidirectional switching behavior results from the *antisymmetric* IEI present in our system.

To reveal the microscopic origins of the necessary effective ISB that is present in our polycrystalline samples, we next measure spatially-resolved magnetic hysteresis loops of the SAFs with the antiparallel coupling by using wide-field Kerr microscopy (see Methods section). In particular, we explore the minor hysteresis loops of the bottom layers (the inset of Fig. 4a) to explore the spatial distribution of the *symmetric* IEI without applying in-plane fields[33]. Interestingly, as shown in Fig. 4a, we find that the average of the two switching fields of the minor loops $H_z^{Avg} = (H_1 + H_2)/2$, which represents the strength of the *symmetric* IEI, has a unidirectional gradient. The axis of this gradient is parallel to the axis **AS** in Fig. 2c: This implies that the effective ISB, another key element for the *antisymmetric* IEI, results from the gradient in the *symmetric* IEI. Given the fact that the *symmetric* IEI is most susceptible to the thickness of the non-magnetic spacers, we speculate that the effective symmetry breaking might be due to a thickness gradient which can naturally appear during the sputtering. Particularly, our sputtered samples are grown without rotation of the sample holder during the growth, therefore, they are likely to lead to such an inhomogeneity, giving rise to the thickness gradient along a certain axis.

Finally, to validate the observed *antisymmetric* IEI and uncover its minimal ingredients, we employ a theoretical ab-initio method to scrutinize this interaction in magnetic heterostructures (Methods section and Supplementary Note 4). We start with a Co/Ru/Pt/Co system with collinear magnetization within each magnetic layer. Here, we take into account the effective in-plane ISB by artificially adjusting in-plane locations of the top Co from hollow sites

"a" and "b" with $C_{3v}$ symmetry into various positions with $C_{1v}$ symmetry as illustrated in Fig. 4b. One of the key manifestations of the *antisymmetric* IEI $\mathbf{D}_{\text{inter}} \cdot (\mathbf{S}_1 \times \mathbf{S}_2)$ is a relativistic contribution to the total energy that is asymmetric with respect to the relative angle $\alpha$ between the magnetic moments $\mathbf{S}_1$ and $\mathbf{S}_2$ in the two Co layers. Indeed, our electronic-structure calculations demonstrate such a unique signature of the *antisymmetric* IEI in the low-symmetric $C_{1v}$ structures (see Fig. 4c and Fig. S7), favoring generally a non-zero canting between adjacent ferromagnetic layers due to the complex interplay with the conventional *symmetric* IEI. To assess the overall relevance of such a chiral *interlayer* interaction, we estimate for comparison the magnitude of the *symmetric* IEI $J_{\text{inter}}(\mathbf{S}_1 \cdot \mathbf{S}_2)$ by using an effective parameter $J_{\text{inter}}$ that describes the small-angle region in the energy dispersion without spin-orbit coupling (SOC). Figure 4d presents the calculated values of both IEIs as a function of the position of the top magnet for an originally ferromagnetic or antiferromagnetic interaction between the ferromagnetic layers. While the *symmetric* interaction exceeds the typical energy scale for the chiral IEI of 1.0 meV by one to two orders of magnitude in the studied system, the latter interaction is more susceptible to changes in the symmetry of the crystal lattice. In particular, the characteristic vector $\mathbf{D}_{\text{inter}}$ is required to be perpendicular to any mirror plane connecting interaction partners in the two layers (see Fig. 4c), which renders the net *antisymmetric* IEI zero in $C_{3v}$ systems but generally finite in the case of reduced symmetry. We would like to point out that, although the microscopic origin of the ISB in our theoretical model is likely not reflecting the full ISB present in the experiment our ab-initio calculations clearly show that the presence of the *antisymmetric* IEI is enabled by any effective ISB. This then favors a chiral magnetization arrangement between separated magnetic layers, thus demonstrating the crucial role of the in-plane ISB for the *antisymmetric* IEI. Any effective in-plane ISB, e.g., from a thickness gradient or a lattice mismatch between different atomic layers, can give rise to the *antisymmetric* IEI. To corroborate this, we demonstrate that for samples grown by an oblique sputtering technique that introduce tilted columnar microstructures[34,35] with a broken inversion symmetry also exhibits the unidirectional and chiral switching behavior indicative of *antisymmetric* IEI. In particular here the **AS** axis is exactly along the sputtering gradient axis (Supplementary Note 5). So our results demonstrate not only that fundamentally an ISB results in an *antisymmetric* IEI but also that the effect can be engineered by tailoring and orienting a thickness gradient for a non-magnetic spacer layer[36] or introducing laterally asymmetric microstructures in a controlled fashion.

So overall, our combined experimental and theoretical work shows that we have completed the set of magnetic exchange interactions in systems with broken inversion symmetry: the *antisymmetric* IEI between two magnetic layers mediated by a non-magnetic spacer results from the ISB that can be tailored to eventually generate and control three-dimensional magnetic textures. Specifically, we experimentally demonstrate the missing component of the IEI in SAFs with parallel and antiparallel alignments, leading to the asymmetric switching behaviors under in-plane bias fields. The observed asymmetric magnetization reversal is a unique signature of the chiral magnetization of the *antisymmetric* IEI. We identify the combination of SOC and the reduced in-plane symmetry as the microscopic origin of the observed antisymmetric IEI. Our findings not only uncover the previously missing *antisymmetric* component of IEI in SAFs with parallel and antiparallel coupling but also opens a path to investigate three-dimensional topological spin structures by *chiral interlayer* interaction that provides the tool for the implementation of three-dimensional topological spin structures in future spintronic devices.

**Acknowledgments**

We acknowledge insightful discussions with Markus Hoffmann, Stefan Blügel, Bertrand Dupé, and Sug-Bong Choe. We acknowledge Fanny Ummelen for private discussions on results that are relevant to this work.[37] D.-S. H, K. L., and M. K. acknowledge support from MaHoJeRo (DAAD Spintronics network, project number 57334897) and the German Research Foundation (in particular SFB TRR 173 Spin+X). K. L. acknowledges the European Union's Horizon 2020 research and innovation programme under the Marie Skłodowska-Curie grant agreement Standard EF No. 709151. C.-Y. Y. acknowledges support from the National Research Foundation of South Korea under Grant 2017R1A2B3002621, and DGIST Research and Development Program under Grant 18-BT-02. M.-H. J. acknowledges support from the National Research Foundation of Korea (NRF) grant funded by the Korea government (MEST) (No. 2017R1A2B3007918). J.-P. H. and Y. M. gratefully acknowledge computing time on the supercomputers JUQUEEN and JURECA at Jülich Super-computing Center, and at the JARA-HPC cluster of RWTH Aachen, as well as funding under SPP 2137 "Skyrmionics" (project MO 1731/7-1) and project MO 1731/5-1 of Deutsche Forschungsgemeinschaft (DFG). K.-W. K. was supported by German Research Foundation (SI 1720/2-1). K.-W. K. and D.-S. H. acknowledge support from the KIST Institutional Program.

**Authors contribution**

M.-H. J. and D.-S. H. conceived the original idea. D.-S. H. and K. L. planned and designed the experiments. D.-S. H. and Y. V. H. fabricated samples with the helps from R. L. and H. J. M. S.. D.-S. H. and K. L. performed transport measurements with W. Y. and data analysis under the supervision of M. K. and M.-H. J.. J.-P. H. and Y. M. performed the first-principle calculations and the analysis of relevant data. K.-W. K. provided theoretical explanations in supplementary note. D.-S. H. and C.-Y. Y. performed the numerical calculation based on a macro-spin model.

D.-S. H. wrote the paper with K. L., J. H., and M. K. All authors discussed the results and commented on the manuscript.

## Methods

**Sample preparation and anomalous Hall measurement.**

The magnetic multilayers were grown on a silicon wafer coated with a 100 nm-thick $SiO_2$ by using a UHV magnetron DC sputtering system at the base pressure of $9.5 \times 10^{-8}$ mbar and the working pressure of $2 \times 10^{-2}$ mbar. Multilayers of Si/Ta(4)/Pt/(4)/Co(1.0)/Pt(0.7)/Ru($t_{Ru}$)/Pt(0.7)/Co(0.9)/Pt(4) were grown at room temperature (layer thicknesses in nm). The Ru is used for the spacer which provides strong IEI, and the Pt layers between top and bottom Co layers are used to enhance the PMA of both ferromagnetic layers. To investigate the Ru-thickness dependent interlayer interaction, a wedge-shaped sample of Ta/Pt/Co/Pt/Ru/Pt/Co/Pt, where the Ru thickness was varied from 0 to 4 nm, was preliminarily grown, and the oscillatory behavior of magnetic hysteresis loops was measured by the magneto-optical Kerr effect in a polar configuration (pMOKE). The Ru thicknesses used in the main text and supplementary notes were selected from the result. The hysteresis loops of the magnetic multilayers were measured by anomalous Hall signal on approximately $5 \times 5$ mm$^2$ sized continuous film by using a Van der Pauw method while sweeping the magnetic field in the out-of-plane and simultaneously applying constant in-plane magnetic field. For the transport measurement, a sinusoidal current with a frequency of 13.7 Hz and a peak-to-peak amplitude of ~1 mA was used as a current source, and a lock-in technique was used for detecting the Hall signal. The spatially resolved magnetic hysteresis loops were measured by using a wide-field Kerr microscope in a polar configuration. The magnetic hysteresis loops at 30 different positions within an approximately $5 \times 5$ mm$^2$ sized continuous film sample of Pt/Co/Pt/Ru/Pt/Co/Pt with antiparallel coupling. The field of view of each area corresponds to ~ $1 \times 1$ mm$^2$. The hysteresis loops are measured sweeping out-of-plane fields with no in-plane fields applied. Thus for all regions, magnetic hysteresis loops symmetric with respect to the out-of-plane field were observed.

**Macro-spin modeling.**

In order to explore the effect of the *antisymmetric* IEI and other magnetic interactions on the magnetization reversal, we employed a macro-spin model that finds an equilibrium magnetization configuration through minimization of the total free energy functional. The total free energy functional consists of anisotropic energy, Zeeman energy, *symmetric* and *antisymmetric* exchange energies, that is given by

$$E_{tot} = -\mu_0 M_{S,top} t_{top} \mathbf{m}_{top} \cdot \mathbf{B} - \mu_0 M_{S,bottom} t_{bottom} \mathbf{m}_{bottom} \cdot \mathbf{B} - K_{top} t_{top} \left(\mathbf{m}_{top} \cdot \hat{z}\right)^2 - K_{bottom} t_{bottom} \left(\mathbf{m}_{bottom} \cdot \hat{z}\right)^2$$
$$- J_{inter} \mathbf{m}_{top} \cdot \mathbf{m}_{bottom} - \mathbf{D}_{inter} \cdot \left(\mathbf{m}_{top} \times \mathbf{m}_{bottom}\right)$$

Here, $M_s$ is saturation magnetization, $\mathbf{m}$ magnetization vector, $K$ effective anisotropy constant, $\mu_0$ vacuum permeability, $\mathbf{B}$ external magnetic field, $t$ thickness of a magnetic layer, $\hat{z}$ unit vector normal to surface, $J_{inter}$ coefficient for *symmetric* IEI, and $\mathbf{D}_{inter}$ DMI vector for *antisymmetric* IEI. The subscript of "top" and "bottom" describe the top and bottom magnetic layers, respectively. For a model system of Pt/Co/Pt/Ru/Pt/Co/Pt, we used the following material parameters: $M_s$= $1.1 \times 10^6$ A/m, $K = 2.24 \times 10^5$ and $5.25 \times 10^5$ J/m$^3$ for the bottom and top layers, respectively. The coefficients for the *symmetric* IEI $J_{inter} = 2.1 \times 10^{-4}$ and $-2.0 \times 10^{-4}$ mJ/m$^2$ and the *antisymmetric* IEI, $D_{inter}$ corresponding to $|D_{inter}/J_{inter}| = 0.1$ and 0.03 were used for the SAFs with parallel and antiparallel coupling, respectively.

**First-principles calculations.**

Using material-specific density functional theory as implemented in the full-potential linearized augmented-plane-wave (FLAPW) code FLEUR,[38] we studied the electronic structure of a thin Co/Ru/Pt/Co film in a super-cell geometry. The lattice constant of the in-plane hexagonal lattice was 5.211 $a_0$ (where $a_0$ is Bohr's radius), the distance between the two Co layers was 12.765 $a_0$, and we assumed a face-centered cubic stacking but variable in-plane positions of the top magnetic layer. Based on the generalized gradient approximation,[39] the self-consistent calculations of the system without SOC were performed using a plane-wave cutoff of 4.0 $a_0^{-1}$, and the full Brillouin zone was sampled by 1024 points. By including the effect of SOC to first order, we unambiguously determined the magnitude of the *antisymmetric* IEI from the change in the energy dispersion of coned spin spirals[40] propagating perpendicular to the film. In these force-theorem calculations with SOC, the Brillouin zone was sampled by 4096 points. Choosing a large enough distance between different super cells, we explicitly ensured that periodic images of the slab do not contribute to the obtained magnetic interaction parameters.

# Figures

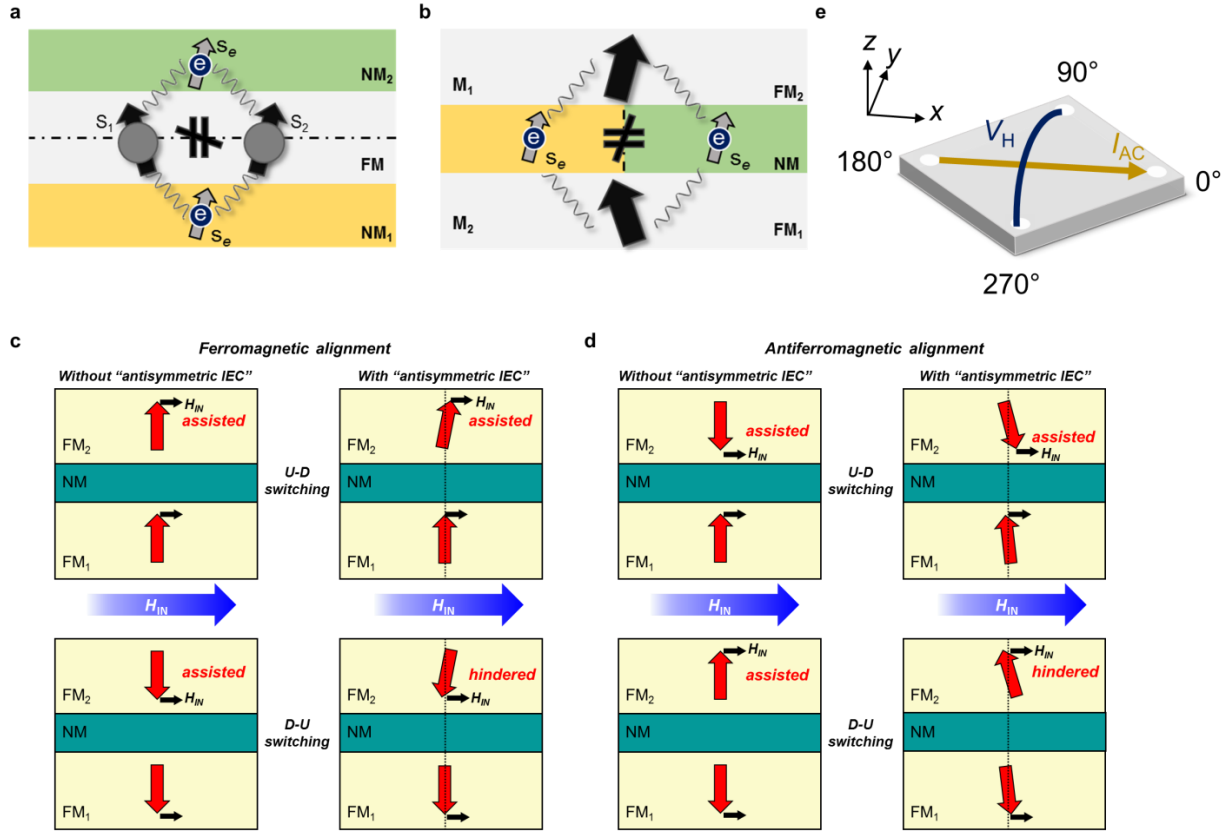

**Figure 1| Schematic illustration of *antisymmetric* exchange interaction and asymmetric switching behavior of perpendicularly magnetized SAFs with parallel and antiparallel coupling by *antisymmetric* IEI.** Schematic illustration of conduction electron-mediated inter-atomic exchange interaction between two atomic spins (**a**) and magnetic layers (**b**), which give rise to *symmetric* and *antisymmetric intralayer* and *interlayer* exchange interactions, respectively. The black arrows in ferromagnets (FMs) represent either localized atomic spins ($S_{1,2}$) or magnetizations ($M_{1,2}$). The gray arrows are spins of conduction electrons in non-magnetic layers (NM). The green and yellow boxes in (**a**) and (**b**) represent a broken inversion symmetry in the out-of-plane and in-plane direction of the films, respectively. All non-magnetic layers are assumed to include heavy elements with SOC. Schematics of symmetric and asymmetric switching of perpendicularly magnetized SAFs with parallel (**c**) and antiparallel alignment (**d**) of the layers due to *symmetric* IEI and additionally in the presence and absence of *antisymmetric* IEI, respectively. The chirality of all magnetization configurations displayed for "with

*antisymmetric* IEI" is right-handed. The red arrows indicate magnetizations of top and bottom FMs. The blue arrows represent an in-plane bias field. The in-plane bias field ($H_{IN}$) breaks the inversion symmetry between up-to-down (U-D) and down-to-up (D-U) switching polarities only in the presence of *antisymmetric* IEI, due to the chiral magnetization alignments. **e** Schematics of the experiments. The AHE is measured in the full sheet samples by using a Van der Pauw method. The ac current is applied parallel to the *x*-axis that is along the 0°-180° line.

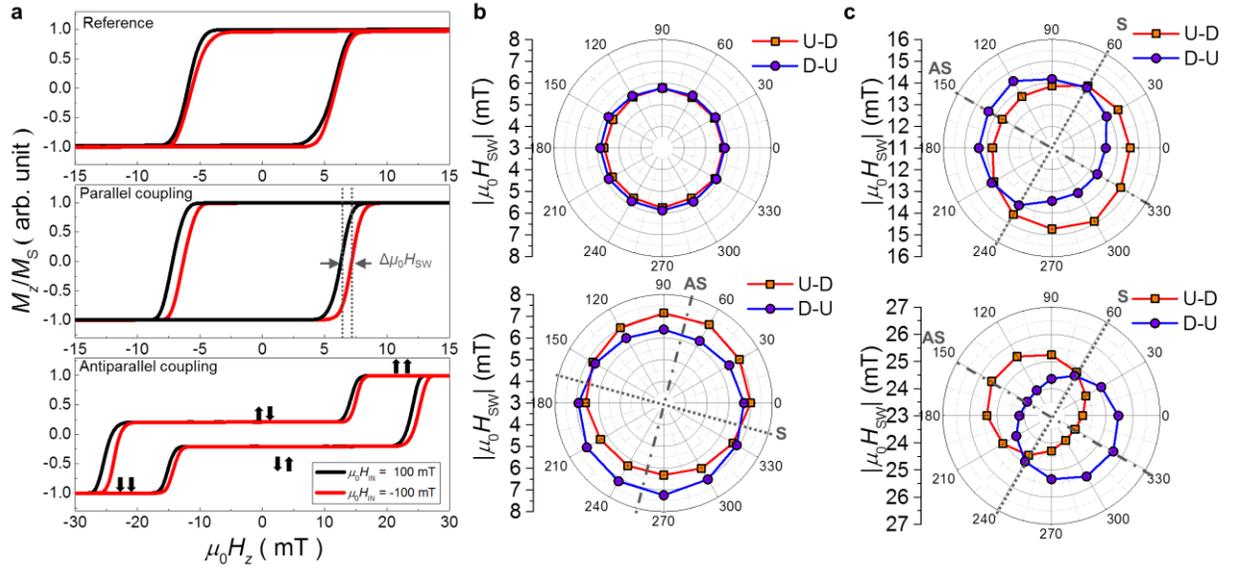

**Figure 2 Chiral and unidirectional magnetization switching behaviors. a** Magnetic hysteresis loops measured by anomalous Hall effect for the reference Pt/Co/Pt/Ru (top panel) and SAFs of Pt/Co/Pt/Ru/Pt/Co/Pt with parallel (middle panel) and antiparallel (bottom panel) coupling. The black and red curves indicate the hysteresis loops under the application of the negative and positive in-plane field of $|\mu_0 H_{IN}|$ = 100mT, respectively, which applied along **AS** axis, as indicated in Fig. 2b and 2c. For the Pt/Co/Pt/Ru/Pt/Co/Pt, the difference in switching fields, $\Delta\mu_0 H_{SW}$ between U-D and D-U corresponds to ~0.7mT. Four representative magnetization configurations which appear during magnetization reversal are indicated as black arrows. **b** Azimuthal-angular dependence of switching field of Pt/Co/Pt/Ru (top panel) and ferromagnetically coupled multilayers of Pt/Co/Pt/Ru/Pt/Co/Pt (bottom panel). The red and blue symbols are for U-D and D-U switching polarities, respectively. The lines are to guide the eyes. **c** Azimuthal-angular dependence of switching field of top (top panel) and bottom (bottom panel) Co layers of antiferromagnetically coupled Pt/Co/Pt/Ru/Pt/Co/Pt. **AS** and **S** represent asymmetric and symmetric axes, respectively.

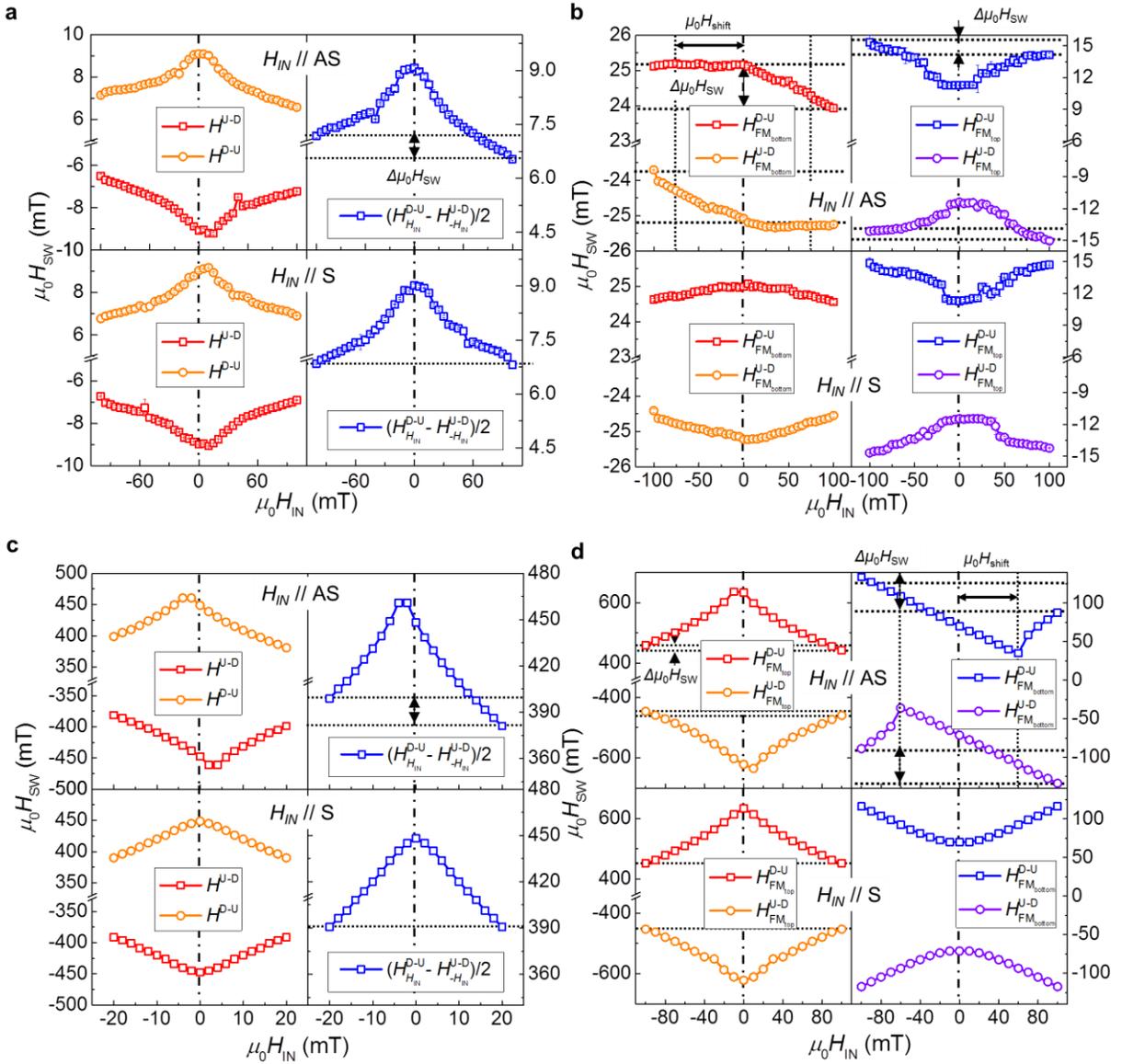

**Figure 3 In-plane field dependence of magnetization switching fields.** Experimentally measured switching field $H_{SW}$ as a function of $H_{IN}$, applied along **AS** (top panel) and **S** (bottom panel) axes as defined in Fig. 2, in SAFs with parallel (**a**) and antiparallel (**b**) coupling. The right panels on each column of (**a**) represent averaged $|H_{SW}|$ of U-D and D-U switching for $H_{IN}$ and $-H_{IN}$, respectively. For both parallel and antiparallel coupled cases, the symmetric (asymmetric) $H_{SW}$ with respect to $H_{IN}=0$ is found when $H_{IN}$ is applied along **S** (**AS**) axis. Calculated $H_{SW}$ as a function of $H_{IN}$ for SAFs with parallel (**c**) and antiparallel (**d**) coupling by using a macro spin model (Methods section).

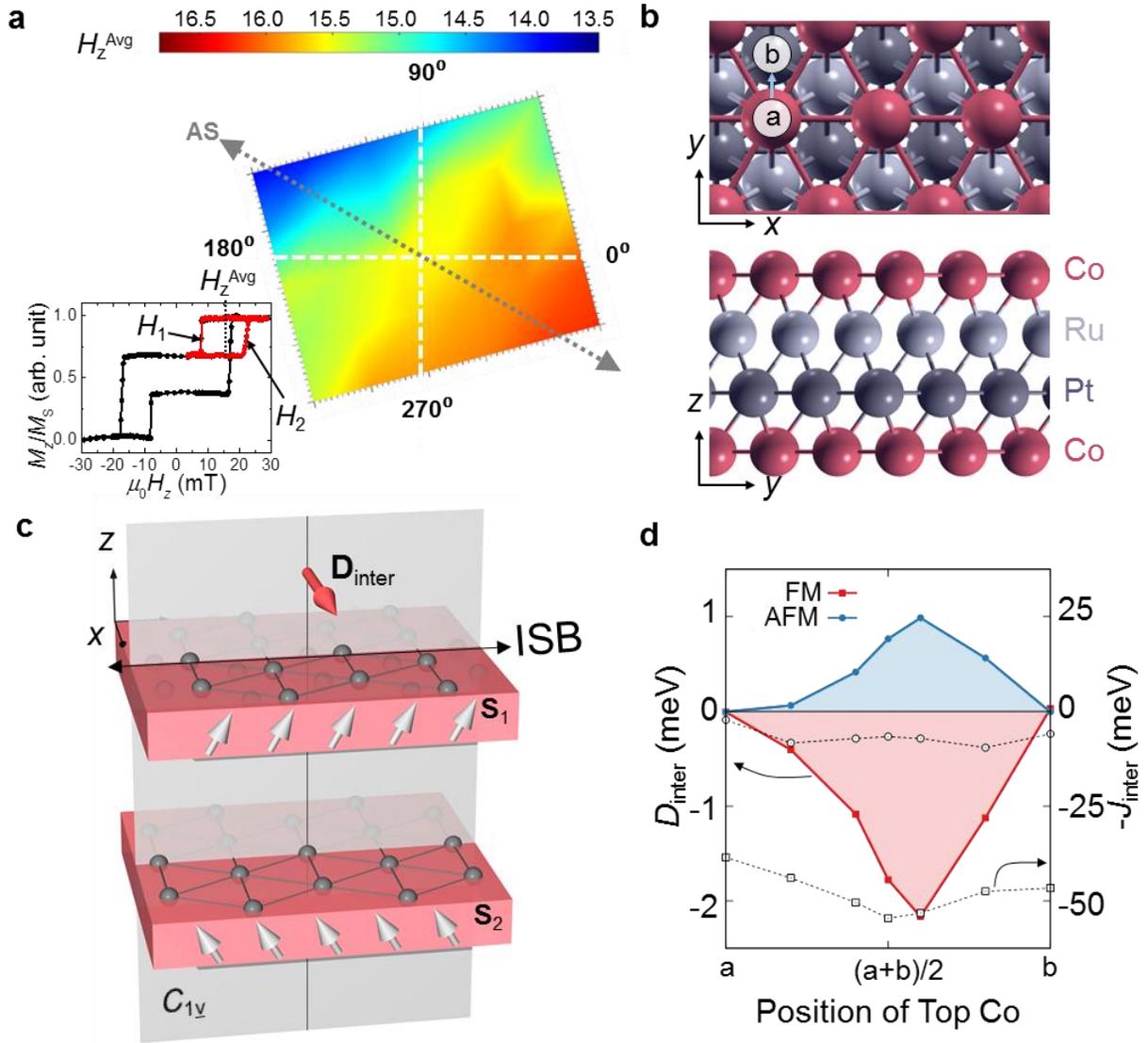

**Figure 4 Effective symmetry breaking in sputtered samples and *antisymmetric* IEI from first principles.**
**a** Spatial distribution of the *symmetric* IEI as obtained by polar Kerr microscopy. The inset represents magnetic hysteresis loops at local area of the sample. The magnetic hysteresis loop is measured without applying an in-plane bias field. The red line indicates a minor loop of the bottom ferromagnetic layer. The $H_z^{Avg}$ is obtained from the average value of two switching fields in the minor loop. The line of 0°-180° corresponds to the current line used for the anomalous Hall measurement. The **AS** axis corresponds to the axis where the unidirectional behavior is the most prominent in Fig. 2. **b** Top and side view of the thin Co/Ru/Pt/Co film. The high-symmetry locations "a" and "b" are marked, and the colored arrow indicates the direction of the considered displacements of the top Co layer. **c** Microscopic schematic of the chiral interlayer exchange in the $C_{1v}$ structures. The collinear magnetization (grey arrows) of adjacent magnetic layers acquires a relative canting due to the *antisymmetric* IEI as mediated by $\boldsymbol{D}_{inter}$, which is perpendicular to the shaded mirror plane. **d** Effective IEI constants $D_{inter}$ (solid lines) and $-J_{inter}$ (dotted lines) as a function of the position of the top Co layer, where squares and circles refer to the cases of parallel and antiparallel coupling, respectively.